\documentclass[twocolumn,showpacs]{revtex4}
\usepackage{graphicx}
\usepackage{bm} 

\begin{document}

\title{Energy spectrum of graphene multilayers in a parallel magnetic field}

\author{Sergey S. Pershoguba and  Victor M. Yakovenko}

\affiliation{Center for Nanophysics and Advanced Materials, Department of Physics, University of Maryland, College Park,
Maryland 20742-4111, USA}

\date{v.3, November 12, 2010}


\begin{abstract}
We study the orbital effect of a strong magnetic field parallel to
the layers on the energy spectrum of the Bernal-stacked graphene
bilayer and multilayers, including graphite.  We consider the
minimal model with the electron tunneling between the nearest
sites in the plane and out of the plane.  Using the semiclassical
analytical approximation and exact numerical diagonalization, we
find that the energy spectrum consists of two domains.  In the
low- and high-energy domains, the semiclassical electron orbits
are closed and open, so the spectra are discrete and continuous,
correspondingly.  The discrete energy levels are the analogs of
the Landau levels for the parallel magnetic field.  They can be
detected experimentally using electron tunneling and optical
spectroscopy.  In both domains, the electron wave functions are
localized on a finite number of graphene layers, so the results
can be applied to graphene multilayers of a finite thickness.
\end{abstract}


\maketitle

\section{Introduction} \label{sec:intro}

Graphene monolayers have attracted much attention recently because of
the unusual Dirac spectrum of electrons \cite{Novoselov-2004}.  A remarkable manifestation of the Dirac dispersion is the unusual spectrum of the Landau levels in a perpendicular magnetic field, resulting in
the anomalous quantum Hall effect (QHE)
\cite{Novoselov-2005,Zhang,Sadowski,Gusynin,Jiang}.  These results stimulated
further investigations of the QHE in the derivatives of graphene.
The unusual Landau levels and the QHE were obtained for a graphene bilayer
in Ref.~\cite{McCann}.  Although the Landau levels in graphite were investigated
a long time ago \cite{Inoue, Dresselhaus,Brandt}, recent studies
\cite{Eva,Plochocka,Orlita,Chuang,Garcia,Koshino} of
graphene multilayers with a moderate number of layers found
interesting features in the Landau spectrum.  Namely, the
spectrum consists of the two families of levels, whose energies scale
as $B$ and $\sqrt{B}$, thus indicating the presence of both
massive and massless Dirac fermions in the system \cite{Eva}.  The
Landau levels for different stacking orders of graphene multilayers
were studied in Ref.~\cite{Guinea}.

On the other hand, much less attention was paid to the orbital effect of a
magnetic field parallel to the graphene layers.  The Shubnikov-de~Haas oscillations were extensively studied in graphite in a tilted magnetic field \cite{Brandt}, but they tend to disappear when the field is parallel to the layers.  Ref.~\cite{Kopelevich} studied the influence of a parallel magnetic field on the putative ferromagnetic, superconducting, and metal-insulator transitions in graphite.  In Ref.~\cite{Iye}, the angular magnetoresistance oscillations (AMRO) were observed in the stage 2 intercalated graphite (in addition to the Shubnikov-de~Haas oscillations) for magnetic fields close to the parallel orientation.  AMRO were first discovered in layered organic conductors \cite{Kartsovnik88,Kajita89} and subsequently observed in many other layered materials: see, e.g., Ref.~\cite{Cooper} and references therein.  Motivated by the experiment \cite{Iye}, a theoretical study of AMRO in graphene multilayers was the original goal of the present paper, with the focus on the peculiarities due to the presence of two sublattice, Dirac spectrum, etc.  However, in the standard theory of AMRO \cite{Cooper}, the interlayer tunneling amplitude is treated as a small perturbation.  It is a reasonable approximation for the intercalated graphite \cite{Dresselhaus-2002}, but not for the pristine graphite, where it is generally accepted that the interlayer tunneling amplitude is quite large.  A non-perturbative treatment of the interlayer tunneling in a tilted magnetic field for graphene multilayers is a very complicated problem.  So, we decided to focus first on the simpler case of the parallel magnetic field.

Additional motivation for this work comes from the recent experiment \cite{Latyshev-2010}, where the current-voltage $I$-$V$ relation was studied for the current perpendicular to the layers in a mesoscopic graphite mesa
consisting of about 20--30 graphene layers.  The experimental
technique is similar to the previous work on the cuprate
superconductors \cite{Krasnov} and the charge-density-wave
materials \cite{Latyshev-2005}.  When a strong parallel magnetic
field up to 55 T is applied to the graphite mesa, the $dI/dV$
curve develops a peak at a non-zero, magnetic-field-dependent
voltage $V$ of the order $80$~mV.  The appearance of the peak may
indicate formation of the Landau levels in the parallel magnetic
field, but detailed interpretation of the experimental results is
currently unclear.

In this paper, we calculate the electron spectrum of two or many coupled
graphene layers in a strong parallel magnetic field.  To simplify the problem, we
consider only the minimal model with the electron tunneling amplitudes
between the nearest sites in the plane ($\gamma_0$) and out of the
plane ($\gamma_1$).  The effect of the higher-order tunneling amplitudes \cite{Gruneis} is briefly discussed in Appendix~\ref{sec:trigonal}.  Our
results should be valid for the energies greater than the energies of
the neglected higher-order tunneling amplitudes and can be verified by tunneling or optical spectroscopy.  We focus only on the orbital effect of the magnetic field and
disregard possible spin effects \cite{Hwang}.  We find some mathematical
similarities between the electron spectrum of graphene multilayers in a
parallel magnetic field and that of quasi-one-dimensional \cite{Yakovenko,Goan}
and quasi-two-dimensional \cite{Lebed} organic conductors \cite{Lebed-book}.

We start with the analysis of a graphene bilayer in a parallel magnetic field
(Sec.~\ref{sec:bilayer}) and then proceed to the infinite number of layers
(Sec.~\ref{sec:graphite}).  We investigate both the quasiclassical electron
orbits in momentum space (Sec.~\ref{sec:trajectories}) and the exact
equation for the energy eigenfunctions, which reduces to the Mathieu
equation (Sec.~\ref{sec:mathieu}).  We employ both the analytical WKB method
and exact numerical diagonalization to find the energy eigenvalues and
eigenfunctions.  We identify the low-energy domain characterized by closed
orbits and discrete spectrum (Sec.~\ref{sec:closed}) and the high-energy domain
with open orbits and continuous spectrum (Sec.~\ref{sec:open}).  The case of a
finite number of layers is analyzed at the end of Sec.~\ref{sec:closed}.  The effect of the tunneling amplitude $\gamma_3$ responsible for trigonal warping is discussed in Appendix \ref{sec:trigonal}.

\section{Graphene bilayer}  \label{sec:bilayer}

\subsection{Model}  \label{sec:bilayer.model}

\begin{figure}
\centering
 \includegraphics[width=0.8\linewidth]{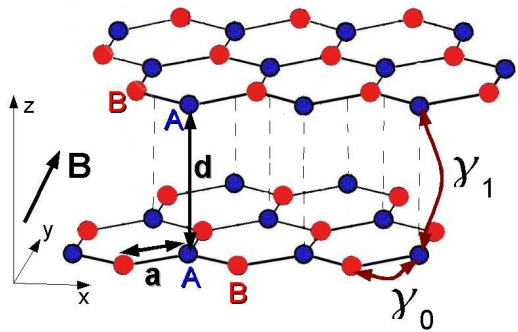}
\caption{(Color online)  A pair of Bernal-stacked graphene layers in the parallel
magnetic field $\bm B$ applied along the $y$ direction. $\gamma_0$
and $\gamma_1$ are the electron tunneling amplitudes.}
 \label{fig:bilayer.geom}
\end{figure}

First, we consider a graphene bilayer and then generalize the
problem to many layers.  The crystal lattice of the Bernal-stacked
graphene bilayer is shown in Fig.~\ref{fig:bilayer.geom}.
The distance between the nearest atoms in graphene is $a=1.4$~\AA, and the
distance between the layers is $d=3.3$~\AA.  We restrict our analysis
to the minimal tight-binding model \cite{Koshino} with the intra-
and inter-layer tunneling amplitudes $\gamma_0 = 3.16$~eV and
$\gamma_1 = 0.38$~eV.

There are two sublattices on each graphene layer.  Thus, the electron
wave function is the vector
\begin{equation}
  \Psi = (\psi_1^A,\psi_1^B,\psi_2^A,\psi_2^B),
\label{4vect}
\end{equation}
where the subscripts 1 and 2 enumerate the layers, and the
superscripts $A$ and $B$ denote sublattices on each layer.
Sublattices can be selected in various ways.  It is convenient for us
to assign the atoms connected by the interlayer tunneling $\gamma_1$
in the Bernal stack to sublattice A and other atoms to sublattice B,
as shown in Fig.~\ref{fig:bilayer.geom}.

In the vicinity of the K point in the Brillouin zone, the electron
Hamiltonian has the form
\begin{equation}
  H =
  \left(
  \begin{array}{cc}
  v_F(\bm p\cdot\bm\sigma) & \gamma_1 I^A \\
  \gamma_1 I^A & v_F (\bm p\cdot\bm\sigma^\ast) \\
  \end{array}
  \right).
\label{bilayer0}
\end{equation}
Hamiltonian (\ref{bilayer0}) acts on the vector (\ref{4vect}).
Correspondingly, $\boldsymbol{\sigma}=(\sigma_x,\sigma_y)$ are the
Pauli matrices acting in the sublattice space; $\bm{p} =
p_x\hat{\bm x}+p_y\hat{\bm y}$  is the in-plane momentum measured
from the K point; $v_F=(3/2\hbar)\gamma_0 a\approx 10^8$~cm/s is
the electron velocity in graphene. The terms $v_F(\bm
p\cdot\bm\sigma)$ and $v_F(\bm p\cdot\bm\sigma^\ast)$ describe the
in-plane Hamiltonians of the graphene layers.  Our choice of the A
and B sublattices results in the diagonal elements having both
$\bm\sigma$ and $\bm\sigma^\ast$ (complex-conjugated) terms. The
term $\gamma_1 I^{A}$ represents the interlayer tunneling, where
the matrix
\begin{equation}
  I^{A} = \frac{1}{2}(I+\sigma_z) =
  \left(
   \begin{array}{cc}
     1 & 0 \\
     0 & 0 \\
   \end{array}
  \right)
\label{IA}
\end{equation}
connects sublattices A of the adjacent graphene layers.

\begin{figure}
\centering
 \includegraphics[width=0.8\linewidth]{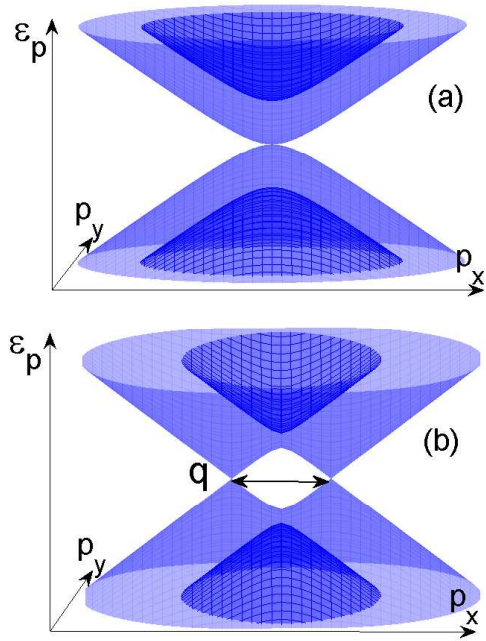}
\caption{(Color online)  (a) The electron spectrum (\ref{regular}) of a graphene bilayer in zero magnetic field. (b) The spectrum (\ref{dispbi}) in a nonzero
parallel magnetic field.  The magnetic field splits the parabolic
spectrum into the two Dirac cones.  An exaggerated value $q=5$ of the
magnetic field parameter was utilized here.}
\label{fig:bilayer.spec}
\end{figure}

Hamiltonian (\ref{bilayer0}) has four eigenvalues
\begin{equation}
  \varepsilon(\bm{p}) =
  \pm\frac{\gamma_1}{2}\pm\sqrt{\frac{\gamma_1^2}{4}+v_F^2p^2}.
\label{regular}
\end{equation}
The well-known spectrum~(\ref{regular}) is shown in
Fig.~\ref{fig:bilayer.spec}(a).  The spectrum consists of the four
bands with the parabolic dispersion for small $p$.

\subsection{Parallel magnetic field}  \label{sec:bilayer.mag.field}

Now let us introduce the in-plane magnetic field $\bm{B}=\hat{\bm
y}B$ applied along the $y$ axis.  We choose the gauge
$\bm{A}=\hat{\bm x}Bz$ and use the Peierls substitution
\begin{equation}
  \bm{p} \rightarrow \bm{p}+\frac{e}{c}\bm{A}.
\end{equation}
Here, we took into account the negative sign of the electron
charge, so $e$ corresponds to its absolute value.  If the layer number
is denoted by $j$, the in-plane electron momentum on the $j$-th
layer changes to
\begin{equation}
  \bm{p}_j = \bm{p} + j\,\Delta p \, \hat{\bm x},
\label{peierls}
\end{equation}
where $\Delta {p}$ is
\begin{equation}
  \Delta {p} = \frac{e}{c}Bd, \qquad
  \frac{\Delta {p}}{\hbar B}=5\times10^{3}\,\rm{cm^{-1} \,T^{-1} }.
\label{Dp}
\end{equation}
The momentum change $\Delta p$ has the following physical meaning.  When an electron
tunnels between the layers, the Lorentz force $\bm{F}=-\frac{e}{c}[\bm{v}\times\bm{B}]$
changes the in-plane momentum by
\begin{equation}
  \Delta p_x=\int F_x dt = \frac{e}{c}B_y \int v_z\,dt = \frac{e}{c}Bd.
\end{equation}
The change in the in-plane momentum results in the relative shift
of the Dirac points on the different layers in the momentum space.

To simplify equations in the rest of the paper, it is convenient
to switch to the dimensionless variables
\begin{eqnarray}
  \frac{v_F\bm p}{\gamma_1} \to \bm p, \quad \frac{v_F\Delta p}{\gamma_1} \to q,
  \quad \frac{\varepsilon}{\gamma_1} \to \varepsilon .
\label{units}
\end{eqnarray}
Here the parameter $q$ is the dimensionless ratio of the ``magnetic shift''
$v_F\Delta p$ and the interlayer tunneling amplitude $\gamma_1$
\begin{equation}
  q = \frac{v_F \Delta p}{\gamma_1}= \frac{v_F}{\gamma_1}
  \left(\frac{e}{c}Bd\right) =0.88\times 10^{-3}B[\rm{T}].
\label{q}
\end{equation}
The parameter $q$ describes the orbital effect of the magnetic field in our model
and will be frequently referred to as the magnetic field for shortness.
It is worth noting that even for a strong magnetic field this
parameter is small $q\ll 1$, e.g., $q=0.044$ for $B=50$~T.

Applying the Peierls substitution (\ref{peierls}) to Hamiltonian
(\ref{bilayer0}) and switching to the dimensionless variables
(\ref{units}), we obtain
\begin{equation}
  H = \left(\begin{array}{cc}
  ((\bm{p}-\bm{q})\cdot\bm{\sigma})  & I^{A} \\
  I^{A} & (\bm{p}\cdot\bm{\sigma}^{\ast}) \\
  \end{array}\right).
\label{bilayer}
\end{equation}
Hamiltonian (\ref{bilayer}) has the following spectrum:
\begin{equation}
  \varepsilon(\bm{p}) =
  \pm{\frac{1}{\sqrt{2}}\sqrt{\bm{p}^2+(\bm{p}-\bm{q})^2+1\pm W}},
  \label{dispbi}
\end{equation}
where
\begin{equation}
  W=\sqrt{\left[(\bm{p}-\bm{q})^2-\bm{p}^2\right]^2+2\bm{p}^2+2(\bm{p}-\bm{q})^2+1}.
\end{equation}
In contrast to  the parabolic dispersion (\ref{regular}), the spectrum (\ref{dispbi}) has
two Dirac points separated by the magnetic shift $q$ with a saddle point in between, as shown in Fig.~\ref{fig:bilayer.spec}(b).  Expanding Eq.~(\ref{dispbi}) around one of the Dirac points
\begin{equation}
  \varepsilon(\bm{p}) \approx \pm\frac{q}{\sqrt{1+q^2}}\,p,
\end{equation}
we find that the slope of the Dirac cones is controlled by the magnetic field and
is greatly reduced for $q\ll1$.

A dispersion similar to Eq.~(\ref{dispbi}) was found for the twisted graphene layers in Ref.~\cite{Santos}, where the relative displacement of the two Dirac cones in the momentum space results from the spatial rotation of the layers
\begin{equation}
  \Delta p_{\rm{rot}}/\hbar = K_0\Delta\phi
  = 6\times10^6\,\rm{cm^{-1}}.
\label{dp.rot}
\end{equation}
Here $\Delta\phi=2^\circ$ is the twist angle, and
$K_0=4\pi/3\sqrt{3}a$ is the distance between the $\Gamma$ and K
points in the reciprocal space.  The saddle point between the two
Dirac points results in the Van Hove singularity in the density of
states, which was observed experimentally in electron
tunneling in Ref.~\cite{Eva1}.  Comparing Eq.~(\ref{Dp}) with
Eq.~(\ref{dp.rot}), we observe that the magnetic field effect is
much weaker than the effect of twisting.  Even for $B=50$~T, the
magnetic shift is $\Delta p/\hbar = 2.5\times10^5\,\rm{cm^{-1}}$
is much smaller than the rotational shift $\Delta
p_{\rm{rot}}/\hbar$.

\section{Graphite}  \label{sec:graphite}

\subsection{Model}  \label{sec:graphite.model}

Now we proceed to the discussion of graphene multilayers.
First we solve the problem for an infinite number of layers, i.e.,\ for
graphite, and then briefly mention the effect of a finite
number of layers.

By analogy with the bilayer Hamiltonian (\ref{bilayer0}), the Hamiltonian of graphite
without magnetic field reads in the adopted units (\ref{units})
\begin{equation}
  H = \left(%
  \begin{array}{ccccc}
     \ddots &  &  &   &   \\
     & (\bm{p}\cdot\bm{\sigma}) & I^{A} &   &   \\
         & I^{A} & (\bm{p}\cdot\bm\sigma^{\ast}) & I^{A} &  \\
         &   & I^{A}  & (\bm{p}\cdot\bm\sigma) &  \\
         &   &   &  & \ddots \\
   \end{array}%
   \right).
\label{hamiltB0a}
\end{equation}
It acts on the vector
\begin{equation}
   \Psi = (\cdots \, \tilde\psi_{j-1} \, \tilde\psi_{j} \, \tilde\psi_{j+1} \cdots),
   \qquad \tilde\psi_j = (\psi_j^A\,\psi_j^B), 
\label{vec}
\end{equation}
where the subscript  $j$ denotes the layer number, and the superscripts
$A$ and $B$ denote sublattices.

\begin{figure}
\centering
 \includegraphics[width=0.85\linewidth]{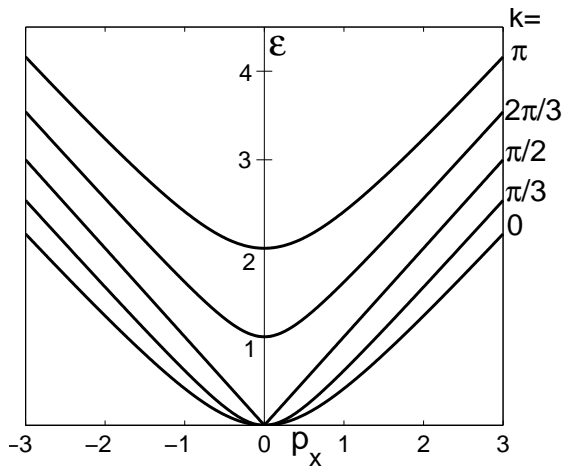}
\caption{Spectrum $\varepsilon_{1}(\bm{p},k)$ (\ref{electrondisp})
of Hamiltonian (\ref{graphite-k-represent}) for $p_y = 0$. Each
curve corresponds to a given value of the out-of-plane momentum
$k$ indicated on the right. The axes are given in the adopted
dimensionless units (\ref{units}).} \label{fig:graphite.spec}.
\end{figure}

Using the momentum representation in the $z$ direction and
introducing the corresponding momentum $k$ (in addition to the
in-plane momentum $\bm{p}$), we transform Hamiltonian
(\ref{hamiltB0a}) into a $4\times4$ matrix similar to the graphene
bilayer Hamiltonian (\ref{bilayer0})
\begin{equation}
  H = \left(\begin{array}{cc}
  (\bm{p}\cdot\bm{\sigma}) &  2 I^{A} \cos k \\
  2I^{A} \cos k & (\bm{p}\cdot\bm{\sigma}^{\ast}) \\
  \end{array}\right).
\label{graphite-k-represent}
\end{equation}
 Hamiltonian~(\ref{graphite-k-represent}) has the following
spectrum
\begin{eqnarray}
  \varepsilon_{1,2}(\bm{p},k) &=& \pm\cos k  +\sqrt{\cos^2k+p^2}, \label{electrondisp}\\
  \varepsilon_{3,4}(\bm{p},k) &=& \pm\cos k - \sqrt{\cos^2k+p^2}.
\label{electrondisp1}
\end{eqnarray}
The subscripts (1,2) and (3,4) denote positive and negative
energies, whereas the subscripts (1,3) and (2,4) correspond to the
terms $\pm\cos k$. Because the spectrum has the electron-hole
symmetry, we consider only the positive energies
$\varepsilon_{1,2}$.  The branches 1 and 2 are equivalent, in the
sense that $\varepsilon_{1}(\bm{p},k+\pi) =
\varepsilon_{2}(\bm{p},k)$.  Thus, it is sufficient to consider
only one branch $\varepsilon_1$, which is plotted in
Fig.~\ref{fig:graphite.spec}.  Since the off-diagonal elements of
Hamiltonian (\ref{graphite-k-represent}) vanish for $k=\pi/2$, the
dispersion has the Dirac-type form $\varepsilon_1(\bm{p},\pi/2)=p$
for $k=\pi/2$, as shown in Fig.~\ref{fig:graphite.spec}.

\subsection{Parallel magnetic field}

\subsubsection{Semiclassical analysis}  \label{sec:trajectories}

\begin{figure}
\centering
  \includegraphics[width=0.9\linewidth]{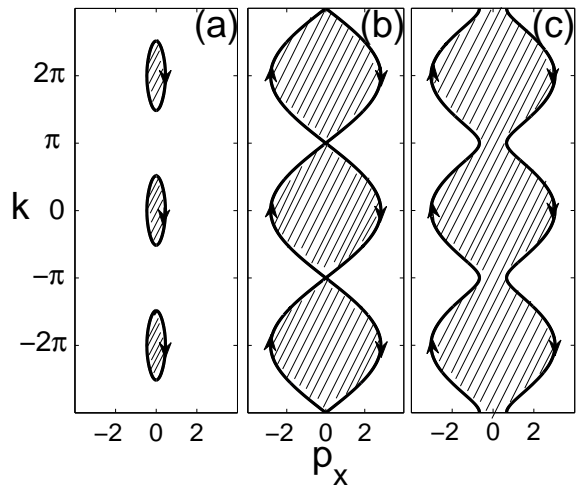}
\caption{Semiclassical electron orbits in the momentum space for
the in-plane magnetic field along the $y$ axis.  Only the orbits
with $p_y=0$ are shown.  They are obtained by intersecting the
$(p_x,k)$ plane with the isoenergetic surfaces
$\varepsilon_1(\bm{p},k)=\varepsilon$ for (a) $\varepsilon=0.1$,
  (b) $\varepsilon=2$,  (c) $\varepsilon=2.2$. The orbits are (a) closed for
 $|\varepsilon|<2$  and (c) open for $|\varepsilon|>2$.}
\label{fig:orbits}
\end{figure}

In the presence of a magnetic field, electrons move along the
isoenergetic surfaces in the momentum space.  For the field
$\bm{B}=\hat{\bm y}B$ along the $y$ direction, the electron orbits
lie on the intersections of the isoenergetic surfaces of the
dispersion~(\ref{electrondisp}) and the planes parallel to the
$(p_x,k)$ plane.  The cross-sections of the isoenergetic surfaces
$\varepsilon_{1}(\bm{p},k)=\varepsilon=\rm {const}$ with the
$(p_x,k)$ plane at $p_y = 0$ are shown in Fig.~\ref{fig:orbits}.
The arrows indicate the direction of electron motion.

Topology of the electron orbits changes with the
increase of the energy $\varepsilon$.  The isoenergetic surfaces for the
dispersion~(\ref{electrondisp}) are closed for $0<\varepsilon<2$, so the
orbits are closed too, see Fig.~\ref{fig:orbits}(a).  Thus,
based on the Onsager quantization rule \cite{Onsager}, the
spectrum is discrete for this energy interval.  However, at the
critical energy $\varepsilon=2$, the isoenergetic surfaces
reconnect, as shown in Fig.~\ref{fig:orbits}(b), and become open for
$\varepsilon>2$, resulting in the open orbits shown in
Fig.~\ref{fig:orbits}(c).  Open semiclassical orbits lead to a continuous
energy spectrum.

\begin{figure}
\centering
  \includegraphics[width=0.85\linewidth]{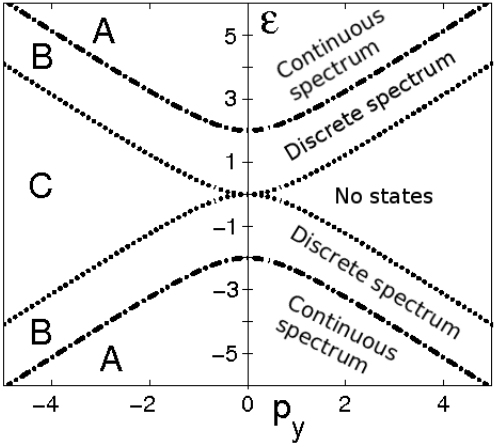}
\caption{Domains of the continuous and discrete spectra in the
($p_y,\varepsilon$) plane.  The dashed-dotted and dotted curves
represent solutions of the equations
$\varepsilon^2-2|\varepsilon|=p_y^2$ and
$\varepsilon^2+2|\varepsilon|=p_y^2$.  The spectrum is continuous
in the region A defined by Eq.~(\ref{crit.cont}), discrete in the
region B defined by Eq.~(\ref{crit.discr}), and there are no
states in the region C defined by Eq.~(\ref{crit.no.states}).}
\label{fig:spec.domains}
\end{figure}

Fig.~\ref{fig:orbits} shows only the electron orbits for $p_y=0$
and $\varepsilon>0$.  We can find the topology of the electron
orbits and the character of the spectrum for an arbitrary $p_y$,
which is a good quantum number for the magnetic field along the
$y$~direction. The orbits are open, so the spectrum is continuous
in $p_x$ for
\begin{equation}
  \varepsilon^2-2|\varepsilon|>p_y^2. \label{crit.cont}
\end{equation}
The orbits are closed, so the spectrum is discrete and
degenerate in $p_x$ for
\begin{equation}
  \varepsilon^2-2|\varepsilon|<p_y^2<\varepsilon^2+2|\varepsilon|.
\label{crit.discr}
\end{equation}
There are no orbits and no states for
\begin{equation}
  p_y^2>\varepsilon^2+2|\varepsilon|.
\label{crit.no.states}
\end{equation}
The domains of the continuous and discrete spectra,
defined by the inequalities (\ref{crit.cont}), (\ref{crit.discr}),
and (\ref{crit.no.states}), are shown in Fig.~\ref{fig:spec.domains}
in the $(p_y,\varepsilon)$ plane.

\subsubsection{Mathieu equation}
\label{sec:mathieu}

Now we present a more formal and exact analysis of the
electron spectrum in a parallel magnetic field.  Applying the Peierls
substitution (\ref{peierls}) in the dimensionless units
\begin{equation}
  \bm{p}_j = \bm{p}+j\bm{q}, \qquad \bm{q}=\hat{\bm x}q
\label{p_j}
\end{equation}
to Hamiltonian (\ref{hamiltB0a}), we obtain
\begin{equation}
H = \left(%
\begin{array}{ccccc}
  \ddots &  &  &   &   \\
   & (\bm{p_{j-1}}\cdot\bm{\sigma}) & I^{A} &   &   \\
         & I^{A} & (\bm{p}_{j}\cdot\bm\sigma^{\ast}) & I^{A} &  \\
         &   & I^{A}  & (\bm{p}_{j+1}\cdot\bm\sigma) &  \\
         &   &   &  & \ddots \\
\end{array}%
\right). \label{hamilt}
\end{equation}
The eigenvalue problem $H\Psi=\varepsilon\Psi$ for $H$ (\ref{hamilt}) and $\Psi$
(\ref{vec}) reads in components
\begin{equation}
  I^{A}(\tilde\psi_{j-1}+\tilde\psi_{j+1})
  +\left[\bm\sigma^{(\ast)}\cdot(\bm{p}+j\bm{q})-\varepsilon\right]\tilde\psi_{j}=0
\label{eigenproblem0}.
\end{equation}
Here, $\bm\sigma^{(\ast)}$ denotes $\bm\sigma$ for even j and
$\bm\sigma^{\ast}$ for odd j.  The matrix equation
(\ref{eigenproblem0}) represents a set of two equations. One of
them relates $\psi_j^B$ and $\psi_j^A$ on the same layer and has
the simple form
\begin{equation}
  \psi_j^B=\frac{p_x\pm ip_y+jq}{\varepsilon}\psi_j^A,
\label{psiB}
\end{equation}
where the signs $\pm$ correspond to even and odd $j$.
Using Eq.~(\ref{psiB}), we algebraically eliminate $\psi_j^B$
components in Eq.~(\ref{eigenproblem0}) and reduce it to the
simpler equation
\begin{equation}
  \psi^A_{j+1}+\psi^A_{j-1} =
  \left(\varepsilon-\frac{(\bm{p}+j\bm{q})^2}{\varepsilon}\right)\psi^A_j,
\label{difference}
\end{equation}
which has the same form for even and odd $j$.  From now on we drop the
superscripts A.  In the Fourier representation
\begin{equation}
  \psi_j = \int_0^{2\pi}\psi(k)\, e^{ikj}dk,
\label{Fourier}
\end{equation}
Eq.~(\ref{difference}) becomes
\begin{equation}
  \left(\frac{d}{dk}-i\frac{p_x}{q}\right)^2\psi(k)-V(k)\,\psi(k)=0,
\label{fullmathieu}
\end{equation}
where
\begin{equation}
  V(k)=\frac{2\varepsilon}{q^2}\cos k
  -\frac{\varepsilon^2-p_y^2}{q^2}.
\label{Vk}
\end{equation}
Here, $\psi(k)$ is a $2\pi$-periodic, twice-differentiable
function $\psi(k)=\psi(k+2\pi)$. To further simplify
Eq.~(\ref{fullmathieu}), we introduce the function $\phi(k)$
\begin{equation}
  \psi(k)=e^{ik(p_x/q)}\phi(k),
\label{psi}
\end{equation}
which eliminates the term $ip_x/q$ from Eq.~(\ref{fullmathieu})
and reduces it to the angular Mathieu equation for $\phi(k)$
\begin{equation}
   \frac{d^2\phi(k)}{dk^2}-V(k)\phi(k) = 0.
\label{mathieu}
\end{equation}
Eq.~(\ref{mathieu}) is equivalent to the Schr\"odinger equation for a particle
moving in the 1D potential $V(k)$ (\ref{Vk}).  The variables $\varepsilon$
and $p_y$ are the parameters that control $V(k)$.

Since $V(k)$ is periodic in $k$, the Bloch theorem can be applied, so the solutions of Eq.~(\ref{mathieu}) have the form
 \begin{equation}
   \phi_{\kappa}(k)=e^{ik\kappa}u_\kappa(k).
 \label{phi}
 \end{equation}
Here, $\kappa$ is the quasimomentum in the space reciprocal to the
$k$ space, and $u_\kappa(k)$ is a $2\pi$-periodic function in $k$.
From Eqs.~(\ref{psi}) and (\ref{phi}) and the periodicity requirement for
$\psi(k)$, we select the solutions of Eq.~(\ref{mathieu}) with $\kappa = -p_x/q$.
Since the solutions of Eq.~(\ref{mathieu}) are periodic in the quasimomentum
$\phi_{\kappa+1}(k)=\phi_{\kappa}(k)$, the parameter $\varepsilon$ in Eq.~(\ref{fullmathieu}) must be periodic in $p_x$: $\varepsilon(p_x)=\varepsilon(p_x+q)$.  Therefore, the magnetic field effectively introduces the magnetic Brillouin zone in $p_x$ with the period $q$.

\begin{figure}
\centering
 \includegraphics[width=0.9\linewidth]{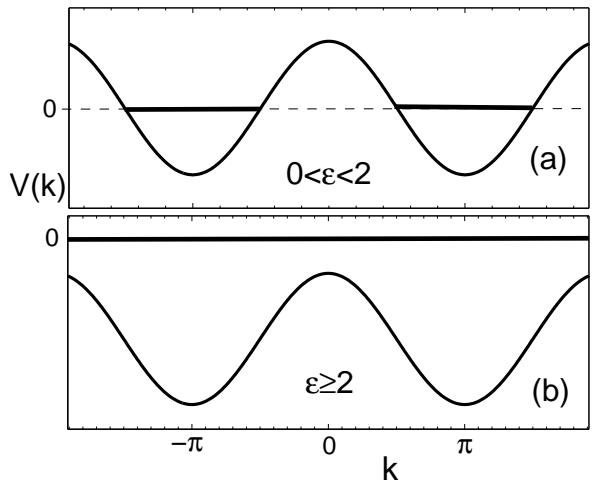}
\caption{The plots of $V(k)$ (\ref{Vk}) for $p_y=0$. The cases of
$\varepsilon=0.1$ and $\varepsilon=2.1$ are shown on the panels
(a) and (b). The classically permitted region
corresponds to $V(k)<0$, as indicated by the thick horizontal
line.  Thus, the panels (a) and (b) represent bounded and unbounded
motion.}
\label{fig:Vk}
\end{figure}

We have reduced the original eigenvalue problem (\ref{difference})
to the convenient differential equation (\ref{mathieu}). Different
regimes for its solutions are controlled by the parameters $p_y$
and $\varepsilon$. If  the criterion~(\ref{crit.discr}) is
satisfied, the 1D classical motion is bounded by the barriers of
$V(k)$, as shown in Fig.~\ref{fig:Vk}(a) for $\varepsilon=0.1$ and
$p_y=0$. Then the energy spectrum is discrete.  On the other hand,
if the criterion~(\ref{crit.cont}) is satisfied, the potential is
negative $V(k)<0$ for any $k$, as shown in Fig.~\ref{fig:Vk}(b)
for $\varepsilon=2.1$ and $p_y=0$.  Then the motion of a particle
is unbounded, and the spectrum is continuous in $p_x$.  The first
regime corresponds to the closed orbits in
Fig.~\ref{fig:orbits}(a), and the second regime to the open orbits
in Fig.~\ref{fig:orbits}(c).  In the following sections, we use
the approaches of both Sec.~\ref{sec:trajectories} and this
section to obtain and interpret the results.

\subsubsection{Closed orbits}
\label{sec:closed}

In this section, we study the electron spectrum in the domain of
the ($p_y,\varepsilon$) plane defined by the criterion
(\ref{crit.discr}) and labeled by the letter B in
Fig.~\ref{fig:spec.domains}.  In this case, the classical motion
of a particle in the 1D potential (\ref{Vk}) is restricted to the
potential wells separated by the barriers, as shown in
Fig.~\ref{fig:Vk}(a).  The height of the barriers is
\begin{equation}
  h(p_y,\varepsilon)={\rm max}[V(k)]=\frac{\varepsilon(2-\varepsilon)+p_y^2}{q^2}.
\end{equation}
The barriers $h(p_y,\varepsilon)\gg1$ are high everywhere, except at
the boundary of the domain~(\ref{crit.discr}).  Thus, we can neglect tunneling and
use the WKB quantization rule for a single well of $V(k)$
\begin{equation}
 4\int\limits_{\arccos{a}}^{\pi}\sqrt{-2\varepsilon\cos k
  +\varepsilon^2-p_y^2}\, dk = 2\pi\left(n+\frac12\right)q,
\label{Onsager}
\end{equation}
where
\begin{equation}
  a = \frac{\varepsilon^2-p_y^2}{2\varepsilon}.
\label{a}
\end{equation}
The integral on the left-hand side of Eq.~(\ref{Onsager}) is
proportional to the area enclosed by the electron orbit in
momentum space, see Fig.~\ref{fig:orbits}(a).  Thus,
Eq.~(\ref{Onsager}) is equivalent to the Onsager quantization rule
in a magnetic field \cite{Onsager}. Using the incomplete elliptic
function of the second kind
\begin{equation}
  E(\phi,m) =
  \int_{0}^{\phi}\sqrt{1-m^2\sin^2\alpha}\,d\alpha,
\label{elliptic}
\end{equation}
Eq.~(\ref{Onsager}) can be written as
\begin{eqnarray}
  && 8\sqrt{2\varepsilon}\sqrt{1+a}\, E\left(\frac{\pi+2\arcsin a}{4},\sqrt{\frac{2}{1+a}}\right)  \nonumber \\
  &&=2\pi \left(n+\frac{1}{2}\right) q.
\label{semiclas}
\end{eqnarray}
Eqs.~(\ref{semiclas}) and (\ref{a}) implicitly define $\varepsilon_n(p_x,p_y)$
as a function of $p_y$ and $n$ for a given magnetic field $q$, and
the spectrum is degenerate in $p_x$.

\begin{figure}
\centering
\begin{tabular}{c}
 \includegraphics[width=0.9\linewidth]{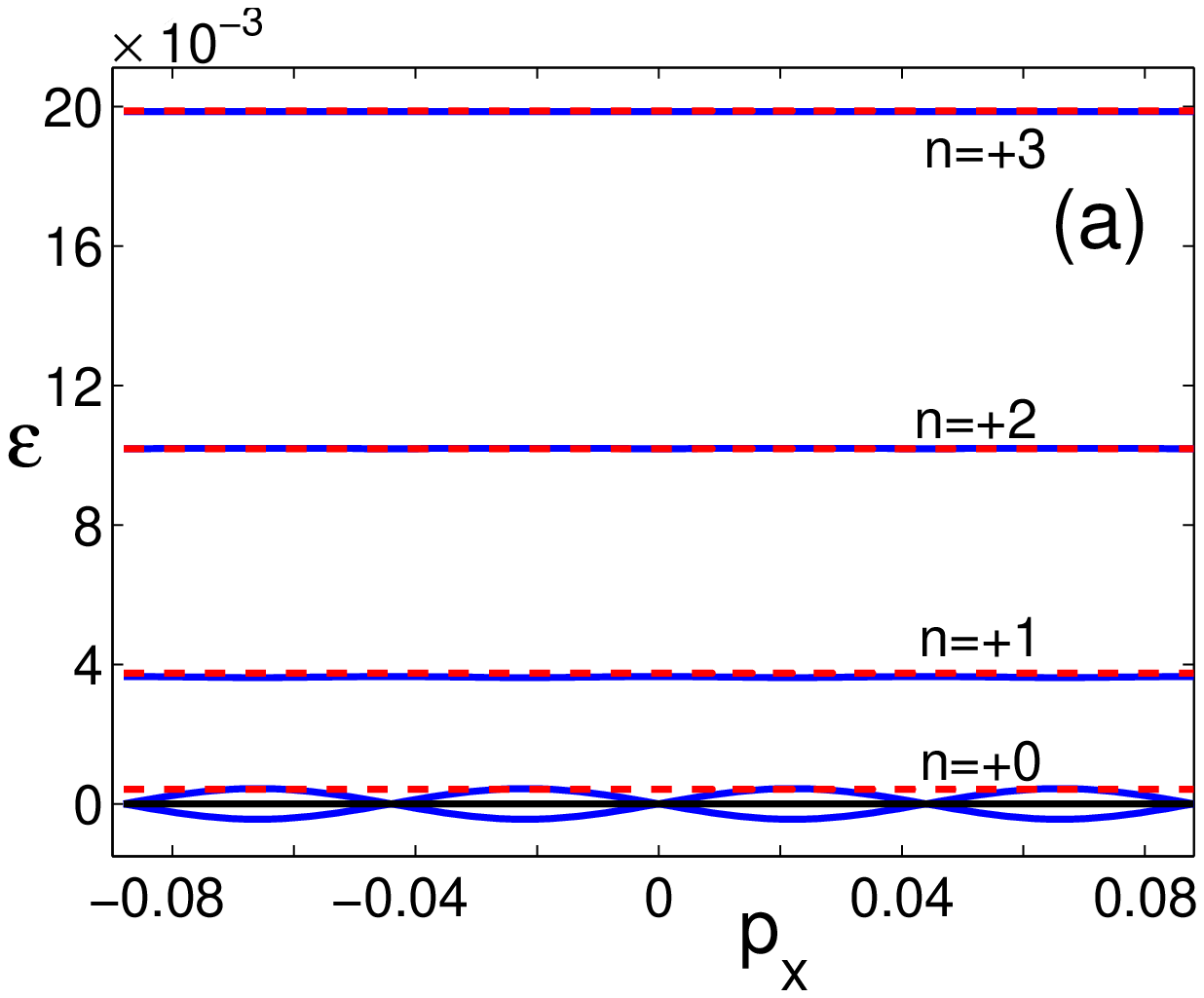} \\
 \includegraphics[width=0.9\linewidth]{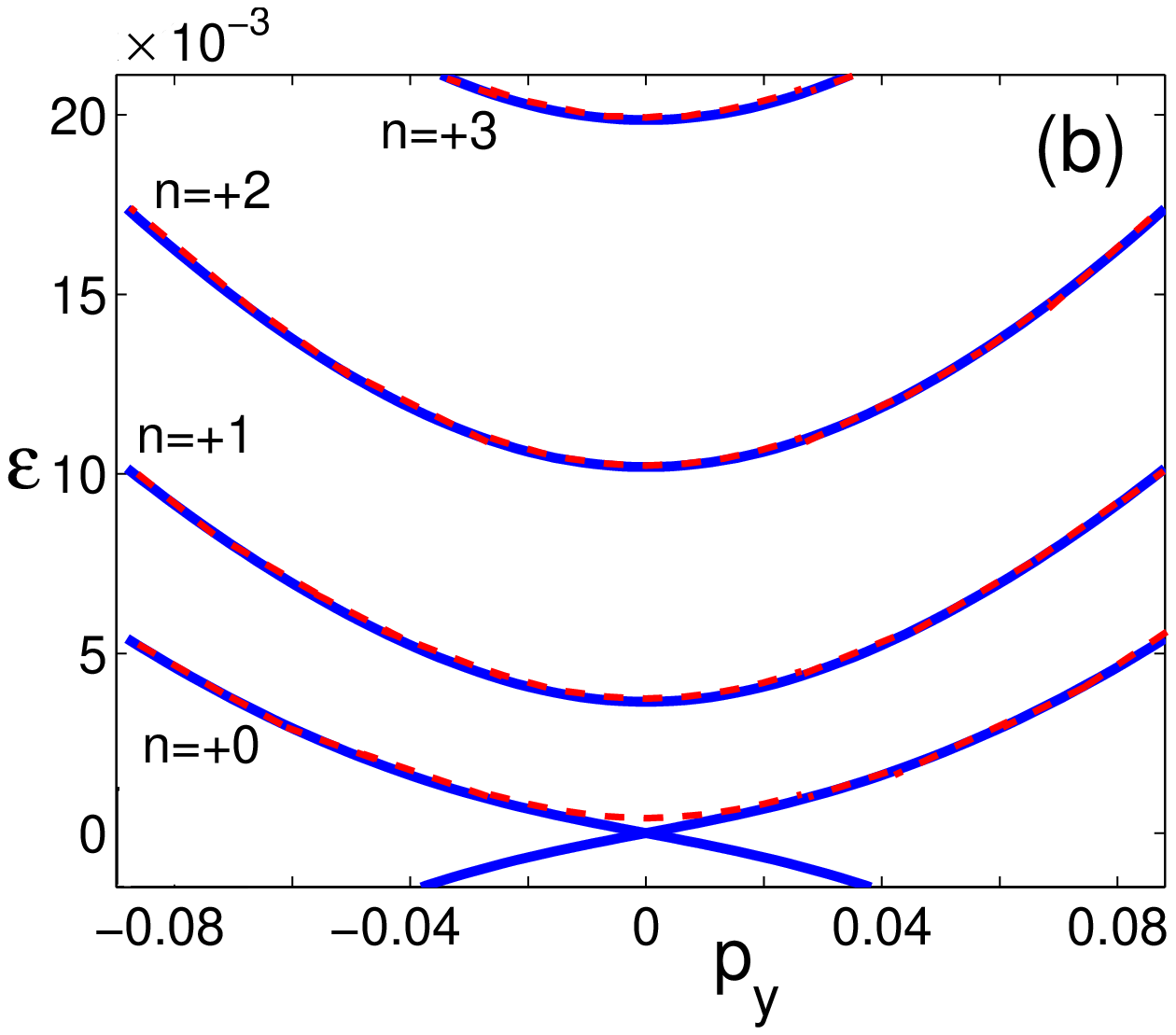} \\
\end{tabular}
\caption{(Color online)  Low-energy levels
$\varepsilon_n(p_x,p_y)$ for $q=0.044$.  Panel (a) shows
$\varepsilon_n(p_x,0)$ vs  $p_x$ for $p_y=0$, and panel (b) shows
$\varepsilon_n(0,p_y)$ vs  $p_y$ for $p_x=0$.  Solid lines
represent exact numerical diagonalization of Hamiltonian
(\ref{hamilt}).  Dashed lines represent the WKB analytical
approximation (\ref{semiclas}).  All quantities are presented in
the adopted dimensionless units (\ref{units}).}
\label{fig:spec.low.levels}
\end{figure}

To check validity of the WKB approximation, we diagonalize of the
original Hamiltonian (\ref{hamilt}) numerically and compare
results with the solutions of Eq.~(\ref{semiclas}).  Momentum
dependences of $\varepsilon_n(p_x,0)$ and $\varepsilon_n(0,p_y)$
are shown in Figs.~\ref{fig:spec.low.levels}(a) and (b) for a few
lowest energy levels at $q=0.044$. The analytical approximation
(\ref{semiclas}) agrees well with the numerical results for $n\neq
0$.  The discrete energy levels shown in
Fig.~\ref{fig:spec.low.levels}(a) are degenerate in $p_x$ and
represent the Landau levels in a parallel magnetic field. However,
the $n=0$ level has a remarkable dispersion in $p_x$.  Similarly
to the spectrum of the graphene bilayer in
Fig.~\ref{fig:bilayer.spec}(b), the $n=0$ level consists of a
series of the Dirac cones shifted by the vector $q$.  This
dispersion cannot be obtained from the approximate WKB equation
(\ref{semiclas}), because of the divergence at $\varepsilon=0$ in
the original equations (\ref{psiB}) and (\ref{difference}).

\begin{figure}
\centering
 \includegraphics[width=0.9\linewidth]{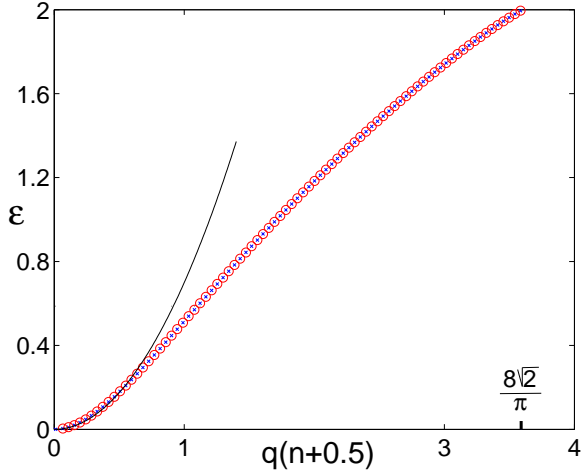}
\caption{(Color online)  Energy levels $\varepsilon_n(0,0)$ vs the
quantum number $n$ at $p_x=0$ and $p_y=0$ for $q=0.044$.  The
horizontal axis shows the combination $q(n+1/2)$.  The circles
represent solutions of the WKB equation (\ref{semiclas}), and the
small points inside the circles represent numerical data.  The
quadratic approximation (\ref{quadratic}) is shown by the solid
line.  All quantities are presented in the adopted dimensionless
units (\ref{units}).} \label{fig:epsilon.vs.n}
\end{figure}

Fig.~\ref{fig:spec.low.levels}(b) shows that the energy levels
$\varepsilon_n(0,p_y)$ have a quadratic dispersion in $p_y$,
except for the $n=0$ level.  Given the degeneracy in $p_x$, the
energy levels $\varepsilon_n(p_x,p_y)$ form one-dimensional bands
in $p_y$, so the density of states diverges at the bottom points
$\varepsilon_n(p_x,0)$ of the bands.  These singularities in the
density of state can be detected experimentally by electron
tunneling or optical spectroscopy. The energies
$\varepsilon_n(0,0)$ are plotted in Fig.~\ref{fig:epsilon.vs.n} in
the interval $0<\varepsilon<2$ vs the combination $q(n+1/2)$
appearing in Eq.~(\ref{semiclas}).  Depending on the magnetic
field $q$, a different number $n_{\rm{max}}$ of the discrete
levels fills the curve.  By setting $\varepsilon=2$ in
Eq.~(\ref{Onsager}), we obtain\begin{equation}
  n_{\mathrm{max}}+\frac{1}{2}=\frac{8\sqrt{2}}{\pi q}=\frac{3.6}{q}.
\end{equation}
For example, for $q=0.044$, we have $n_{\rm{max}}=81$ levels,
which are depicted by circles in Fig.~\ref{fig:epsilon.vs.n}.

\begin{figure}
\includegraphics[width=0.9\linewidth]{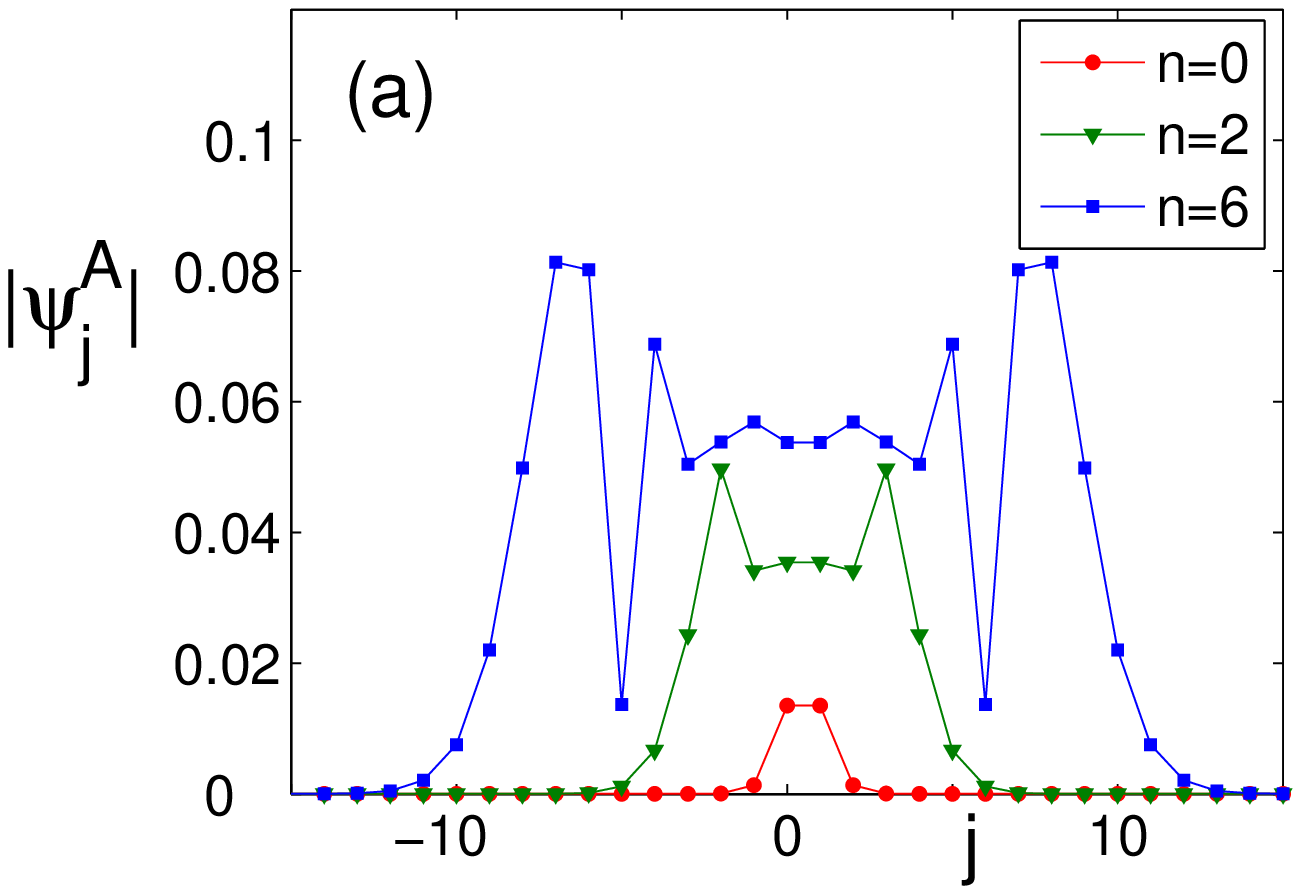} \\
\includegraphics[width=0.9\linewidth]{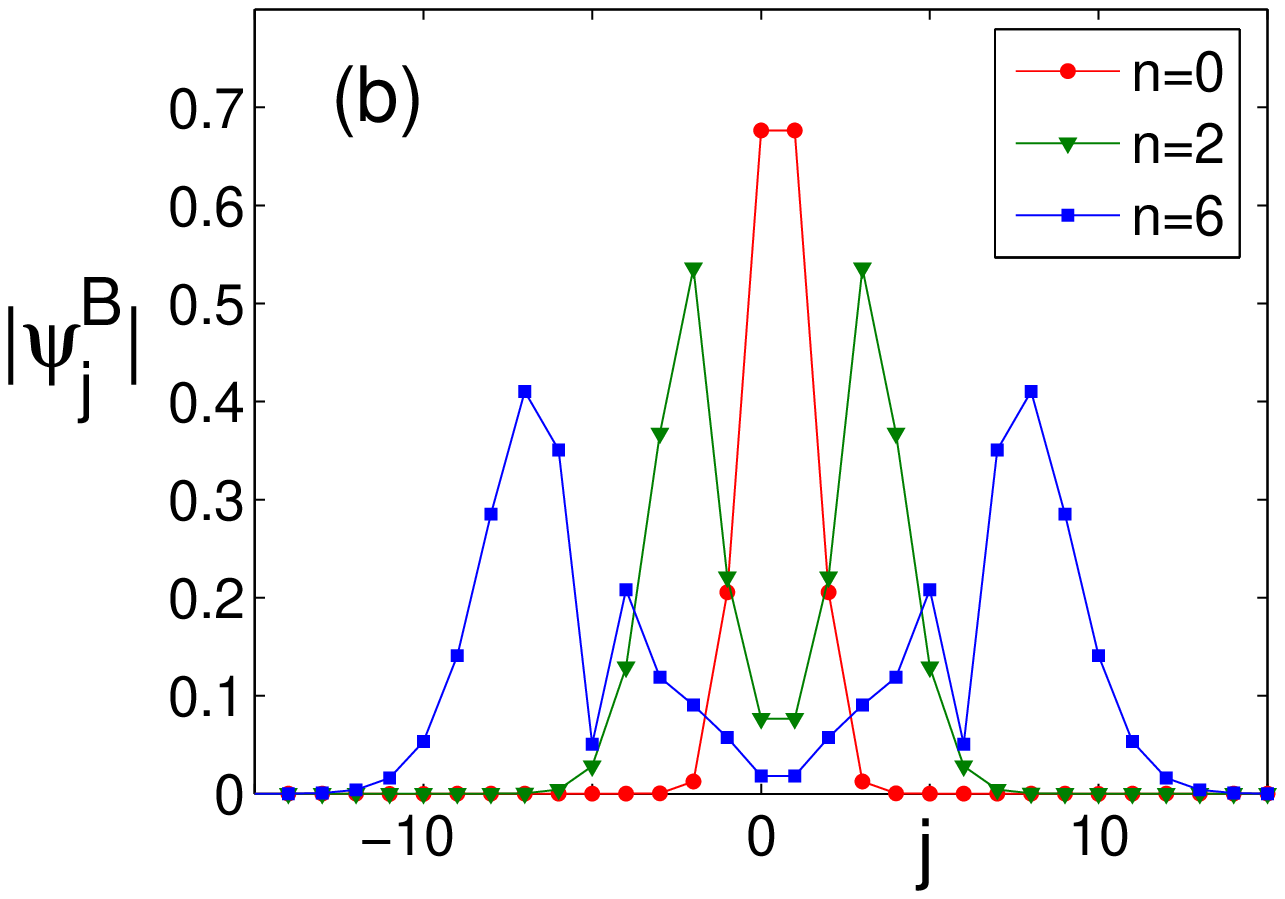}
\caption{(Color online)  The absolute values of the wave functions
$|\psi_j^A|$ and $|\psi_j^B|$ on the sublattices A and B vs the
layer number $j$ for $q=0.044$, $p_x=q/2$, and $p_y=0$.  The
variable $n$ denotes the energy level number.} \label{fig:psi.j}
\end{figure}

For small $\varepsilon$ and $p_y=0$, we find from Eq.~(\ref{semiclas})
\begin{equation}
 \varepsilon_n=\frac{\pi^2}{32E^2\left(\frac\pi4,\sqrt{2}\right)}
 q^2\left(n+\frac{1}{2}\right)^2 = 0.7\,q^2\left(n+\frac{1}{2}\right)^2.
\label{quadratic}
\end{equation}
We observe that the energies (\ref{quadratic}) depend
quadratically on the level number $n$ and the magnetic field $q$.
This dependence is different from the usual Landau level
dependence, where the energies are linear in the field and in the
quantum number.  The reason for the unusual dependence in our case
is the following.  For small $\varepsilon$, the semiclassical orbit
in Fig.~\ref{fig:orbits}(a) shrinks to a thin ellipsoid of the length
$\pi$ in the $k$~direction and the width $\sqrt{2\varepsilon}$ in
the $p_x$ direction.  Thus, the area enclosed by the semiclassical
orbit is proportional to $\sqrt{\varepsilon}$, so the Onsager
quantization rule gives quadratic dependence of the energy on the
magnetic field and the level number $n$.  The quadratic
approximation (\ref{quadratic}) is shown by the solid line in
Fig.~\ref{fig:epsilon.vs.n} and works well in the region
$\varepsilon<0.1$.

Fig.~\ref{fig:psi.j} shows the plots of $|\psi_j^A|$ and
$|\psi_j^B|$ vs the layer number $j$ for several energy levels
$n$.  We observe that the magnetic field causes localization of
the wave function on a finite number of layers.  According to
Eq.~(\ref{psiB}), the wave functions for the low energy levels
$\varepsilon_n$ are localized predominantly on the sublattice B.
The magnitudes of $|\psi_j^A|$ and $|\psi_j^B|$ are shown by
different vertical scales in panels (a) and (b) of
Fig.~\ref{fig:psi.j}.

\begin{figure}
\centering
\includegraphics[width=0.9\linewidth]{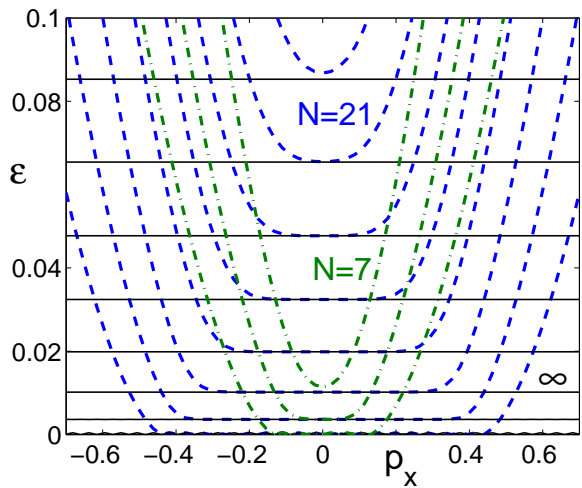}
\caption{(Color online)  Energy spectrum $\varepsilon_n(p_x,0)$ vs
$p_x$ at $p_y=0$ and $q=0.044$ for the system with a finite number
of layers $N$.  The dashed-dotted, dashed, and solid lines
correspond to $N=7$, 21, and $\infty$.  All quantities are
presented in the adopted dimensionless units (\ref{units}).}
\label{fig:spec.finite.n}
\end{figure}

Now let us briefly discuss the spectrum of a finite system with
the total number of layers $N$.  Fig.~\ref{fig:spec.finite.n} shows
$\varepsilon_n(p_x,0)$ for $N=7$, 21, and $\infty$.  The degeneracy in $p_x$
is lifted for a finite number of layers, but, with increasing $N$,
the spectrum approaches to that of the infinite system with $N=\infty$.
Indeed, if the localization length for a particular energy level is shorter
than the size of the system, the energy of the level is the same as for
$N=\infty$.  Thus, the results obtained for $N=\infty$ are applicable to a finite system with a sufficient large $N$.

\subsubsection{Open orbits}
\label{sec:open}

Now we study the energy spectrum in the domain defined by
Eq.~(\ref{crit.cont}) and labeled by the letter A in
Fig.~\ref{fig:spec.domains}.  It corresponds to the open electron
orbits in Fig.~\ref{fig:orbits}(c).  In the Mathieu
equation (\ref{mathieu}), the potential $V(k)<0$ is negative for
any $k$, so the motion is unbounded, as shown in Fig.~\ref{fig:Vk}(b).
Then, the WKB solutions of Eq.~(\ref{mathieu}) are
\begin{equation}
  \phi(k) = e^{\pm i S(k)}, \qquad
  S(k) = \int_{0}^{k}\sqrt{|V(k)|}\,dk,
\label{wkb}
\end{equation}
where the signs $\pm$ correspond to the direction of motion.  Because of
the periodicity requirement for $\psi(k)$ and Eq.~(\ref{psi}), the phase
accumulation in Eq.~(\ref{wkb}) over the period $2\pi$ must be equal
to $-2\pi p_x/q$ plus an integer multiple of $2\pi$.  Thus we obtain the
following quantization condition for the open orbits
\begin{equation}
  q S(2\pi)= 2\int\limits_0^{\pi}\sqrt{-2\varepsilon\cos
  k+\varepsilon^2-p_y^2}\,dk
  =\mp 2\pi(p_x+\tilde{n}q).
\label{Onsager0}
\end{equation}
Here the integer $\tilde{n}$ is different from the integer $n$ in
Eq.~(\ref{Onsager}) and takes both negative and positive values.
The sign of $p_x+\tilde{n}q$ corresponds to the two solutions in
Eq.~(\ref{wkb}).  Eq.~(\ref{Onsager0}) can be represented in terms
of the elliptic integral (\ref{elliptic})
\begin{equation}
  4\sqrt{2\varepsilon}\sqrt{1+a}\,\,
  E\left(\frac{\pi}{2},\sqrt{\frac{2}{1+a}}\right) =
  \mp 2\pi(p_x+\tilde{n}q), \label{semiclasO}
\end{equation}
where the parameter $a$ is given by Eq.~(\ref{a}).  In contrast to
Eq.~(\ref{semiclas}), Eq.~(\ref{semiclasO}) contains $p_x$, so the
energy levels $\varepsilon_n(p_x,p_y)$ continuously depend on $p_x$.

\begin{figure}
\centering
\includegraphics[width=0.9\linewidth]{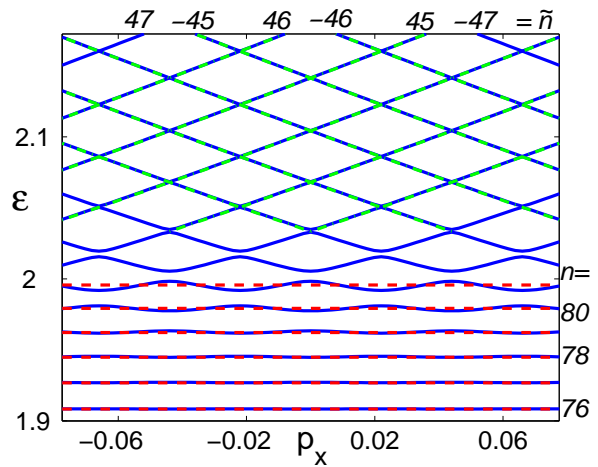}
\caption{(Color online)  Energy spectrum around $\varepsilon=2$
for $q=0.044$ and $p_y=0$.  The solid lines are obtained by
numerical diagonalization of Hamiltonian (\ref{hamilt}).  The
dashed lines, obtained from Eq.~(\ref{semiclas}), are labeled by
the integer $n$ shown on the right.  The dashed-dotted lines,
obtained from Eq.~(\ref{semiclasO}), are labeled by the integer
$\tilde{n}$ shown at the top.  This plot illustrates a transition
from discrete to continuous spectrum.  All quantities are
presented in the dimensionless units (\ref{units}).}
\label{fig:spec.transition}
\end{figure}

The energy spectra obtained from Eqs.~(\ref{semiclas}) and
(\ref{semiclasO}) are compared in Fig.~\ref{fig:spec.transition}
with the results of numerical diagonalization of
Hamiltonian~(\ref{hamilt}) around the critical energy
$\varepsilon=2$ for $p_y=0$.  For $\varepsilon<2$, the spectrum
consists of the energy levels degenerate in $p_x$, which are well
described by the analytical approximation (\ref{semiclas}) for
closed orbits.  The corresponding level number $n$ is shown on the
right in Fig.~\ref{fig:spec.transition}. At the energy
$\varepsilon=2$, the spectrum undergoes a transition to the regime
of continuous dispersion in $p_x$.  For $\varepsilon>2$, the
spectrum consists of the two families of lines with the opposite
slopes.  This spectrum is well described by the analytical
approximation (\ref{semiclasO}) for open orbits.  The
corresponding number $\tilde{n}$ labels the dispersion lines and
takes both positive and negative values shown at the top in
Fig.~\ref{fig:spec.transition}. Because the left-hand sides of
Eqs.~(\ref{semiclas}) and (\ref{semiclasO}) differ by the factor
of 2 at $\varepsilon=2$ and $p_x=0$, the numbers $n$ and
$\tilde{n}$ are connected as $n\approx2|\tilde{n}|$.  The
approximations (\ref{semiclas}) and (\ref{semiclasO}) stop working
in the vicinity of the critical energy $\varepsilon=2$.

For a high energy $\varepsilon$, when the parameter $a$ (\ref{a}) is
large, we can obtain the spectrum explicitly by expanding the square root
in Eq.~(\ref{wkb}) for $S(k)$ in powers of $1/a$
\begin{equation}
  S(k) = \frac{\sqrt{\varepsilon^2-p_y^2}}{q}k-\frac{\varepsilon
  \sin k}{q\sqrt{\varepsilon^2-p_y^2}}.
\label{s.approx}
\end{equation}
Then the quantization condition (\ref{Onsager0}) gives
\begin{equation}
  \sqrt{\varepsilon^2-p_y^2} = \mp (p_x+\tilde{n}q),
\label{subst}
\end{equation}
which can be written as
\begin{equation}
  \varepsilon_{\tilde{n}}^2 = (p_x+\tilde{n}q)^2+p_y^2.
\label{spec.Dirac.decoupled}
\end{equation}
The spectrum Eq.~(\ref{spec.Dirac.decoupled}) is the same as for
decoupled graphene layers with the Peierls substitution (\ref{p_j}).

\begin{figure}
\centering
\includegraphics[width=0.9\linewidth]{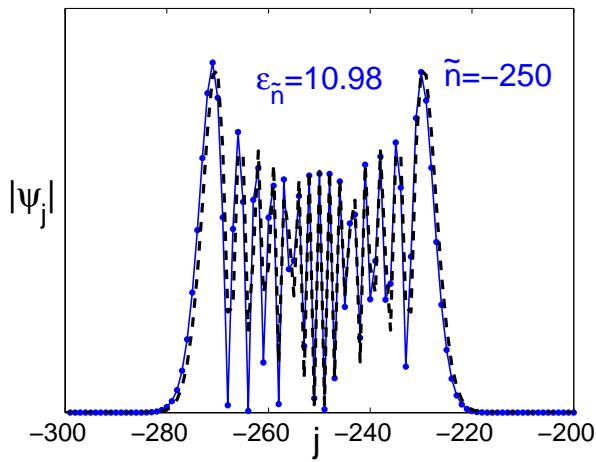}
\caption{(Color online)  The absolute value of the wave function
$|\psi_j|$ vs the layer number $j$ for a state from the domain of
continuous spectrum.  The parameters of the plot are
$\tilde{n}=-250$, $\varepsilon_{\tilde{n}} =10.98$, $p_x=p_y=0$,
and $q=0.044$. The solid and dashed lines represent exact
numerical diagonalization of Hamiltonian (\ref{hamilt}) and the
approximate analytical formula (\ref{psi.j.simple}),
respectively.} \label{fig:psi.j.open}
\end{figure}

Substituting the square root expression from Eq.~(\ref{subst})
into Eq.~(\ref{s.approx}), we obtain the approximate WKB wave
functions (\ref{wkb}) for $\varepsilon>0$
\begin{equation}
  \phi(k)= \exp\left( -i\frac{p_x+\tilde{n}q}{q}k+i\frac{\sqrt{(p_x+\tilde{n}q)^2+p_y^2}}{q(p_x+\tilde{n}q)}\sin
  k\right).
\end{equation}
Then, using Eq.~(\ref{psi}), we calculate the Fourier transform
(\ref{Fourier}) and find the wave function $\psi_j$ in the
direct space
\begin{equation}
  \psi_j =
  J_{\tilde{n}-j}\left(\frac{\sqrt{(p_x+\tilde{n}q)^2+p_y^2}}{q(p_x+\tilde{n}q)}\right).
\label{psi.j}
\end{equation}
Here $J_m(x)$ is the Bessel function of the $m$-th order of the first
kind. For $p_y=0$, Eq.~(\ref{psi.j}) simplifies to
\begin{equation}
  \psi_j = J_{\tilde{n}-j}\left(\frac{\mathrm{sign}(p_x+\tilde{n}q)}{q}\right).
\label{psi.j.simple}
\end{equation}
The wave function (\ref{psi.j.simple}) is centered at $j=\tilde
n$, as shown in Fig.~\ref{fig:psi.j.open}.  We observe that, even
though Eq.~(\ref{spec.Dirac.decoupled}) coincides with the
spectrum of effectively decoupled graphene layers, the
corresponding wave function (\ref{psi.j.simple}) is localized on a
large number of layers proportional to $1/q$.  Similar wave
functions are known for the quasi-one-dimensional conductors in a
magnetic field \cite{Yakovenko,Goan}.

\section{Conclusions}

In this work, we have studied the orbital effect of a strong
magnetic field applied in the $y$ direction parallel to the layers
of the graphene bilayer and multilayers. For the former, the
magnetic field splits the parabolic bilayer dispersion into the
two Dirac cones in the momentum space with the spacing
proportional to the magnetic field.  For the latter, we have found
two domains in the parameter space with distinct energy spectra.
In the low-energy domain, the semiclassical electron orbits are
closed, so the spectrum is discrete and degenerate in $p_x$.  The
energy levels depends quadratically on $p_y$, thus forming a
series of one-dimensional bands in $p_y$.  The discrete energies
of the bottoms of the bands are the analogs of the Landau levels
but depend quadratically on the energy level number $n$ and the
magnetic field $B$.  The $n=0$ energy level around zero energy has
unusual properties and consists of a series of shifted Dirac
cones, similarly to the bilayer case.  In the high-energy domain,
the semiclassical electron orbits are open, so the spectrum in
continuous in $p_x$, thus forming two-dimensional bands in $p_x$
and $p_y$.  For high enough energies, these bands evolve into the
Dirac cones originating from different layers and shifted in the
momentum space due to the applied magnetic field.  In both
regimes, the wave functions are localized on a finite number of
layers.  Mathematically, the problem reduces to the Mathieu
equation.  The WKB approximation for the semiclassical electron
orbits in the momentum space in the magnetic field agrees well
with exact numerical diagonalization, except for a few special
cases, where the WKB approach is not applicable.

The obtained energy spectrum can be verified experimentally using electron tunneling or optical spectroscopy.  Our results may help to understand the experimentally measured $I$-$V$ curves for a mesoscopic graphite mesa in a strong parallel magnetic field up to 55 T \cite{Latyshev-2010}, although detailed interpretation is not clear at this point.

We studied the minimal model with the two tunneling amplitudes between the nearest neighboring sites in the plane ($\gamma_0$) and out of the plane ($\gamma_1$).  In general, the obtained results should be valid for the energies greater than the neglected higher-order tunneling amplitudes \cite{Gruneis}, which can be taken into account in future studies, if necessary.  A more detailed discussion of the influence of the trigonal warping amplitude $\gamma_3$ on our results is given in Appendix \ref{sec:trigonal}.

\begin{acknowledgements}
The authors thank Paco Guinea and Eva Andrei for discussions and
Yuri Latyshev for sharing results of Ref.~\cite{Latyshev-2010}.
V.M.Y.\ is grateful to KITP for hospitality at the program on
Low-Dimensional Electron Systems in April--June 2009, where this
work was initiated.
\end{acknowledgements}

\appendix

\section{The effect of the higher-order tunneling amplitudes} \label{sec:trigonal}

In this appendix, we examine how our results are affected by
inclusion of the higher-order tunneling amplitudes \cite{Gruneis}.
We focus on the tunneling amplitude $\gamma_3=0.29$~eV
\cite{Gruneis}, which connects any given atom B with the three
nearest atoms B on the adjacent layers (see
Fig.~\ref{fig:bilayer.geom}).  This term has the C$_3$ rotational
symmetry and is responsible for the trigonal warping of the
electron spectrum.  Given that the amplitudes $\gamma_3$ and
$\gamma_1$ are of the same order, we need to explain why one can
disregard $\gamma_3$, but use $\gamma_1$ at the same time.

In the presence of $\gamma_3$, the model Hamiltonian becomes
\begin{eqnarray}
  H =  \left(
  \begin{array}{cccc}
  0 & v_F p & 2\gamma_1\cos k  & 0 \\
  {\rm c.c.} & 0 & 0 & 2\gamma_3'v_F p\cos k \\
  {\rm c.c.} & 0 & 0 & v_F p^\ast \\
  0 & {\rm c.c.} & {\rm c.c.} & 0 \\
  \end{array}
  \right),
\label{H2}
\end{eqnarray}
where c.c.\ means complex conjugated, and we introduced
$p=p_x+ip_y$ and the small dimensionless parameter
\begin{equation}
  \gamma_3'\equiv\frac{\gamma_3}{\gamma_0}=0.08.
\label{gamma3}
\end{equation}
Because the $\gamma_3$ term vanishes at the K point ($p=0$), we expanded this term in Eq.~(\ref{H2}) to the first order in $p$, to be consistent with the linearization of the $\gamma_0$ term.  In contrast, the $\gamma_1$ term does not vanish at the K point.  Thus, the $\gamma_3$ term in Eq.~(\ref{H2}) is smaller than the $\gamma_1$ term by the small factor $\gamma_3'$ for the energies where the linearization in $p$ is applicable, even though $\gamma_1$ and $\gamma_3$ are of the same order.
For this reason, the $\gamma_3$ term can be generally neglected relative to the $\gamma_1$ term, except for the very low energies, as discussed below.

Hamiltonian (\ref{H2}) is written in the conventional system of
units.  In the system of units (\ref{units}), adopted in our work,
Hamiltonian becomes
\begin{eqnarray}
  H =  \left(%
  \begin{array}{cccc}
  0 &  p & 2\cos k  & 0 \\
  p^\ast & 0 & 0 & 2\gamma_3' p\cos k \\
  2\cos k & 0 & 0 & p^\ast \\
  0 & 2\gamma_3' p^\ast\cos k &  p & 0 \\
  \end{array}%
  \right).
\label{H3}
\end{eqnarray}
Hamiltonian (\ref{H3}) differs from Hamiltonian
(\ref{graphite-k-represent}) by the additional terms proportional
to $\gamma_3'$.

\begin{figure}[b]
 \includegraphics[width=0.49\linewidth]{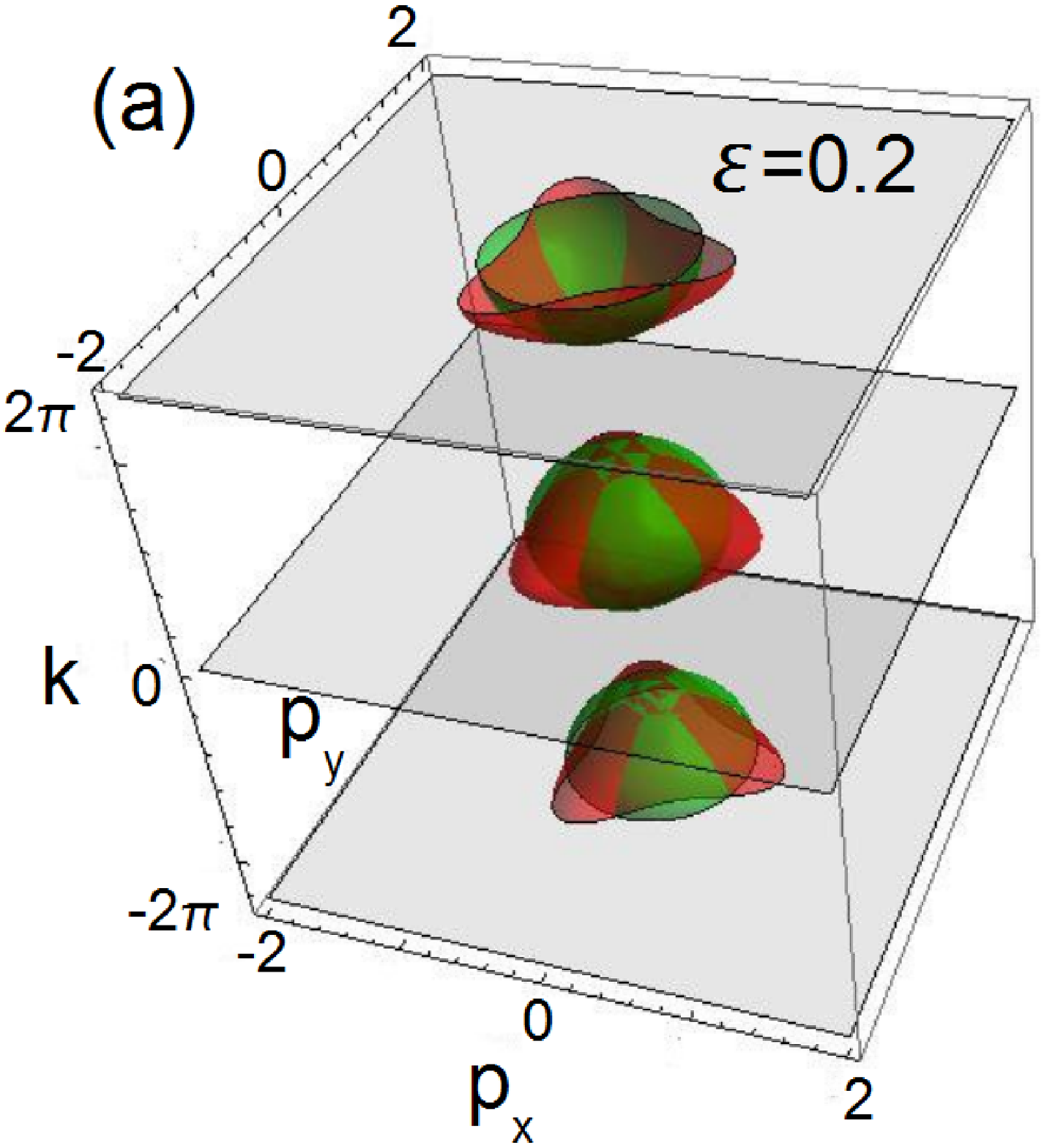} \hfill
 \includegraphics[width=0.49\linewidth]{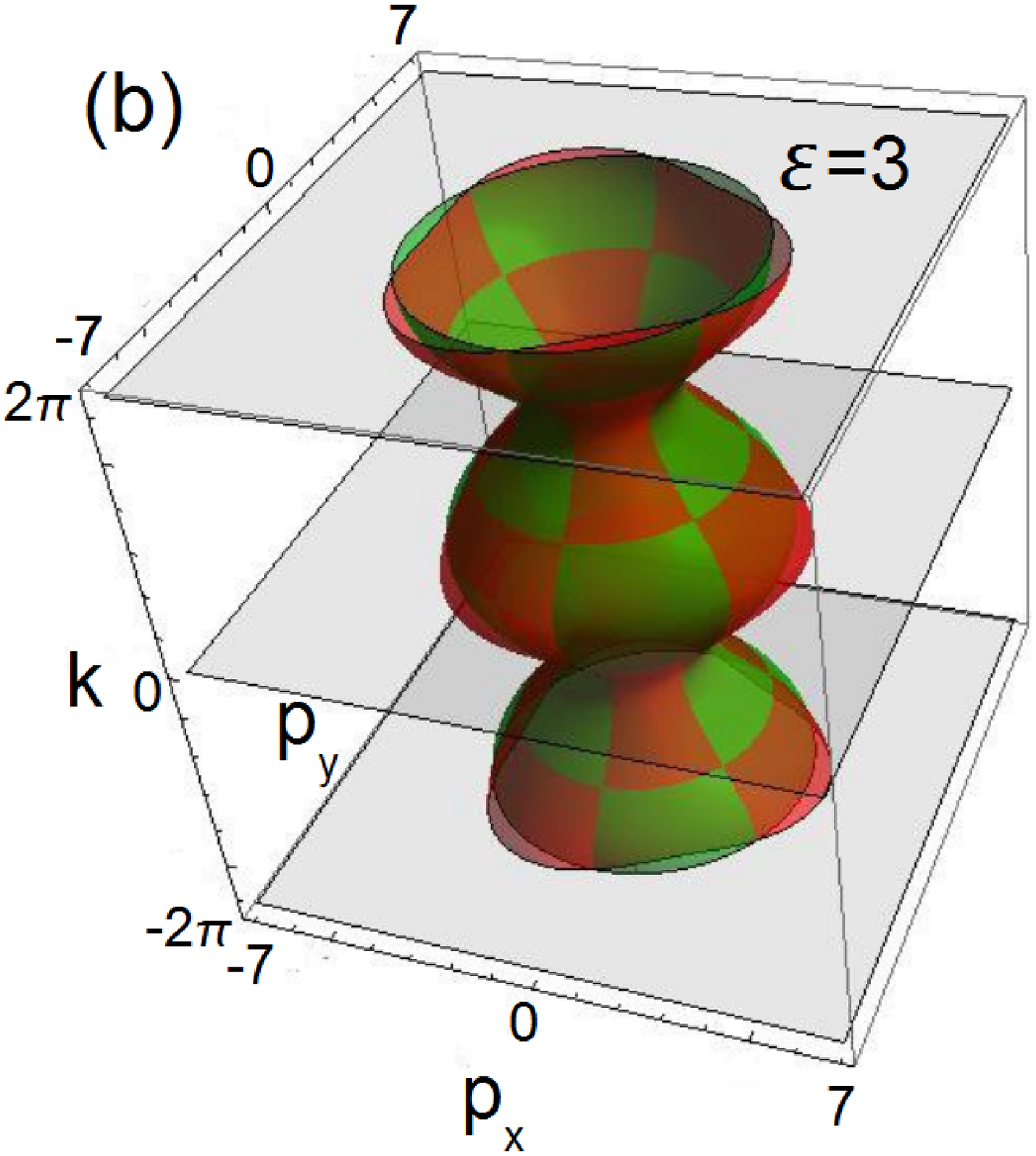}
\caption{(Color online)  Isoenergetic surfaces
$\varepsilon(p_x,p_y,k)=\varepsilon=\rm const$ are shown for
$\varepsilon=0.2$ and $\varepsilon=3$ in panels (a) and (b),
respectively.  The green cylindrically-symmetric surfaces
correspond to Hamiltonian (\ref{graphite-k-represent}), and the
red trigonally-warped surfaces to Hamiltonian (\ref{H3}).}
\label{fig:surfaces}
\end{figure}

Our results for the electron spectrum in a parallel magnetic field
rely mainly on the shape and topology of the isoenergetic surfaces
$\varepsilon(\bm p,k)=\varepsilon=\rm const$ in the momentum
space.  The isoenergetic surfaces for Hamiltonians (\ref{H3}) and
(\ref{graphite-k-represent}) are compared in
Fig.~\ref{fig:surfaces} for $\varepsilon=0.2$ in panel (a) and for
$\varepsilon=3$ in panel (b).  The green, cylindrically-symmetric
surfaces correspond to Hamiltonian (\ref{graphite-k-represent})
(their cross sections are shown in Fig.~\ref{fig:orbits}), whereas
the red surfaces with the C$_3$ symmetry correspond to
Hamiltonian~(\ref{H3}).  We observe that $\gamma_3$ produces
trigonal warping of the isoenergetic surfaces, but the topology of
the surfaces does not change for $|\varepsilon|>0.1$.  Thus, our
conclusions about the discrete energy spectrum for
$|\varepsilon|<2$ and continuous for $|\varepsilon|>2$ in the
presence of a parallel magnetic field remain qualitatively valid.
However, the energy spectrum acquires anisotropy with respect to
the in-plane rotation of the magnetic field, which can be studied
experimentally.

Trigonal warping of the isoenergetic surfaces becomes progressively more pronounced at low energies $|\varepsilon|<0.1$.  At much lower energies $\varepsilon\sim 0.01$, the isoenergetic surface spits into four separate branches
with three new Dirac points surrounding the original Dirac point
\cite{Brandt,Gruneis}.  For such low energies, the isoenergetic surfaces
qualitatively differ from the case of $\gamma_3=0$, and our results are not applicable.  At the low energies, it is also necessary to take into account the other tunneling amplitudes, in particular $\gamma_2\sim10$~meV, which connects the next-nearest layers \cite{Gruneis}.  In the presence of many tunneling amplitudes, the problem becomes very complicated.  In addition, disorder may smear out the spectrum at low energies.

Thus, we restrict the applicability of our results to the relatively high energies
$|\varepsilon|>0.1$, which is about 40~meV in the dimensional units.  The energy spectrum in this range can be studied by tunneling or optical spectroscopy.  Notice that the magnetic-field-dependent peak in $dI/dV$ was observed in Ref.~\cite{Latyshev-2010} at the applied voltage of about 80 meV, although exact nature of this peak is still unclear.



\end{document}